\documentclass[useAMS,usenatbib]{mn2e}
\usepackage{graphicx,txfonts}

\defcitealias{PalmerB_83a}{PE83}
\defcitealias{LovisC_07a}{LP07}
\defcitealias{NorlenG_73a}{N73}
\defcitealias{WhalingW_95a}{W95}
\defcitealias{WhalingW_02a}{W02}
\defcitealias{DeCuyperJ-P_98a}{DCH98}

\newcommand{\ms}{\hbox{${\rm m\,s}^{-1}$}}
\newcommand{\kms}{\hbox{${\rm km\,s}^{-1}$}}

\newcommand{\apropto}{\,\rlap{\raise 0.4ex\hbox{$\propto$}}{\lower 0.5ex\hbox{$\sim$}}\,}

\newcommand{\bspsmall}{\vspace{0.5cm}\small\noindent This paper has been typeset
from a \TeX/\LaTeX\ file prepared by the author.\normalsize}

\title[ThAr wavelength calibration lines]
{Selection of ThAr lines for wavelength calibration of echelle spectra
  and implications for variations in the fine-structure constant}
\author[M.~T.~Murphy et al.]{
  M.~T.~Murphy$^{1}$\thanks{E-mail: mim@ast.cam.ac.uk (MTM)}, P.~Tzanavaris$^{2,3}$, J.~K.~Webb$^{2}$, C.~Lovis$^{4}$\\
  $^{1}$Institute of Astronomy, University of Cambridge, Madingley
  Road, Cambridge, CB3 0HA, UK\\
  $^{2}$School of Physics, University of New South Wales, Sydney
  N.S.W.~2052, Australia\\
  $^{3}$Institute of Astronomy and Astrophysics, National Observatory
  of Athens, I.~Metaxa \& V.~Pavlou, 152 36 Penteli, Greece\\
  $^{4}$Observatoire de Gen\`{e}ve, 51 Ch.~des Maillettes, 1290 Sauverny, Switzerland
}

\begin{document}

\date{Accepted 2007 March 20. Received 2007 March 15; in original form 2006 November 29}

\pagerange{\pageref{firstpage}--\pageref{lastpage}} \pubyear{2007}

\maketitle

\label{firstpage}

\begin{abstract}
  Echelle spectrographs currently provide some of the most precise and
  detailed spectra in astronomy, the interpretation of which sometimes
  depends on the wavelength calibration accuracy. In some
  applications, such as constraining cosmological variations in the
  fundamental constants from quasar absorption lines, the wavelength
  calibration is crucial. Here we detail an algorithm for selecting
  thorium-argon (ThAr) emission lines for wavelength calibration which
  incorporates the properties of both a new laboratory wavelength list
  and the spectrograph of interest. We apply the algorithm to the Very
  Large Telescope Ultraviolet and Visual Echelle Spectrograph
  (UVES) and demonstrate a factor of $\ga$3 improvement in the
  wavelength calibration residuals (i.e.~random errors) alone. It is
  also found that UVES spectra calibrated using a previous, widely
  distributed line-list contain systematic $\pm30$--$75{\rm
    \,m\,s}^{-1}$ distortions of the wavelength scale over both short
  and long wavelength ranges. These distortions have important
  implications for current UVES constraints on cosmological variations
  in the fine-structure constant. The induced systematic errors are
  most severe for Mg/Fe{\sc \,ii} quasar absorbers in the redshift
  range $1.2\!\la\!z_{\rm abs}\!\la\!2.3$, with individual absorbers
  studied by recent authors containing systematic errors up to 4 times
  larger than quoted statistical errors.
\end{abstract}

\begin{keywords}
  atomic data -- line: identification -- techniques: spectroscopic --
  atlases -- quasars: absorption lines
\end{keywords}


\section{Introduction}\label{sec:intro}

Echelle spectrographs are the most common instrument for recording
high-dispersion astronomical spectra. Since many echelle diffraction
orders can be cross-dispersed and thereby recorded simultaneously on
rectangular-format media [e.g.~modern charge-coupled devices (CCDs)],
more-or-less continuous wavelength coverage over much of the optical
range can be achieved whilst maintaining high resolving
power\footnote{Resolving power $R \equiv \lambda/{\rm FWHM}$, where
  ${\rm FWHM}$ is the instrumental profile's full-width at
  half-maximum.}. To fully exploit the velocity precision available in
high-resolution spectra, precise wavelength calibration over much of
the optical range is required. This is now usually done by comparison
with exposures of a thorium-argon (ThAr) hollow-cathode emission-line
lamp \citep*[e.g.][]{BreckinridgeJ_75a}.

After \citet{MeggersW_55a} first suggested that Th transitions might
act as reliable wavelength standards, many authors
\citep[e.g.][]{GiacchettiA_70a} measured Th laboratory wavelengths and
confirmed their suitability for such purposes. There are several
reasons the ThAr spectrum has proven an enduring standard for
astronomical spectroscopy: (i) Most optical ThAr transitions have
measured laboratory wavelengths, most of which are resistant to
pressure and Stark shifts; (ii) Th has one stable isotope and
terrestrial Ar is almost entirely composed of $^{40}$Ar
\citep{RosmanK_98a}. Also, both Th and $^{40}$Ar have even nuclei.
Therefore, no unresolved isotopic or hyperfine structure in the ThAr
lines is expected; (iii) Most ThAr lines are unresolved at most
practical resolving powers, i.e.~$R\la150000$; (iv) The ThAr
line-density is high, $\ga$1 line per \AA, over most of the optical
range; (v) ThAr lamps are fairly cheap, are commercially available and
have lifetimes of typically $\ga$1\,yr. Other calibration sources
include iodine absorption cells \citep[e.g.][]{MarcyG_92a} but these
degrade the object signal and can only be used over a comparatively
short wavelength interval (typically $\sim$5000--6500\,\AA). Another,
potentially ultra-stable, absolute calibration system might be based
on the new technology of laser frequency combs \citep{MurphyM_07e}, a
possibility now under development.

Despite the distinct advantages, there are, however, several drawbacks
to using ThAr spectra for wavelength calibration. Firstly, different
ThAr lines have widely different intensities, far beyond the typical
dynamic range of modern CCDs (i.e.~a factor of $\sim$50000). Secondly,
the wavelength distribution of lines is highly non-uniform; it turns
out that {\it most} ThAr lines are blended with relatively strong
nearby lines, even at $R\la150000$.  It is therefore important to
carefully select which ThAr lines are to be used when calibrating
echelle spectra. This task is complicated by the fact that a given
line's intensity (relative to other lines) varies from lamp to lamp
and even within the one lamp over time.  ThAr lines arise from a
variety of Th and Ar ionization states and so, as the lamp ages and/or
the operating current and pressure are varied, the relative strengths
of different lines can change. Thus, ideally, detailed selection of
ThAr lines for wavelength calibration is best carried out separately
for each ThAr exposure.  However, in practice, most
spectrograph-specific data reduction pipelines carry out only simple
and minimal line selection, thereby assuming that a detailed
pre-selection of lines has already been carried out. That is, for most
practical purposes, it suffices to have a ThAr line-list pre-selected
for the spectrograph of interest.

In this paper we detail an algorithm to pre-select ThAr lines for any
echelle spectrograph and we apply the algorithm to the Ultraviolet and
Visual Echelle Spectrograph (UVES) on the European Southern
Observatory's (ESO's) Very Large Telescope (VLT) on Cerro Paranal,
Chile. ESO have already pre-selected ThAr lines to calibrate UVES and
the line-list is included with the ESO pipeline for UVES data
reduction. Their selection derives from the results of
\citet[hereafter \citetalias{DeCuyperJ-P_98a}]{DeCuyperJ-P_98a} who
constructed simulated ThAr emission spectra based on line tables in
the literature. \citetalias{DeCuyperJ-P_98a} report `effective'
wavelengths for selected ThAr lines given the expected degree of line
blending based on the relative line intensities in the literature. The
effective wavelengths therefore depend on resolution; the ESO list is
based on the $R=100000$ results of
\citetalias{DeCuyperJ-P_98a}\footnote{http://www.eso.org/instruments/uves/tools/tharatlas.html}.
However, the ESO list is not identical to the
\citetalias{DeCuyperJ-P_98a} list and seems to include $>500$ lines
which were rejected by \citetalias{DeCuyperJ-P_98a} because no
effective wavelengths could be found. Furthermore, all wavelengths are
only given to 3 decimal places (in units of \AA) and, in many cases,
the wavelengths were {\it truncated}, rather than rounded, from 4
decimal places.

Such treatment of the ThAr line-list can cause non-negligible
systematic effects in several applications. In this paper we focus on
the effects it has had on current constraints on cosmological
variations in the fine-structure constant, $\alpha_{\rm em}$, derived
from VLT/UVES quasar absorption spectra. If $\alpha_{\rm em}$ were
different in the high-redshift absorption clouds, the relative
wavelengths of different metal-line transitions would differ from
those observed in laboratories on Earth. This effect provides a
precise observational probe of possible variations in $\alpha_{\rm
  em}$ over cosmological time- and distance-scales called the
`many-multiplet' (MM) method \citep{DzubaV_99a,WebbJ_99a}. Analyses of
a large sample of quasar spectra from the Keck High Resolution Echelle
Spectrograph (HIRES) have so far provided the first evidence for
variation in $\alpha_{\rm em}$
\citep{WebbJ_99a,WebbJ_01a,MurphyM_01a,MurphyM_03a} with the most
recent results indicating a smaller $\alpha_{\rm em}$ in the
absorption clouds at the fractional level of
$\Delta\alpha/\alpha=(-0.57\pm0.11)\times10^{-5}$ \citep{MurphyM_04a}.
Clearly, such a potentially fundamental result must be refuted or
confirmed with many different telescopes and spectrographs. Some first
attempts with VLT/UVES have now been made
\citep[e.g.][]{ChandH_04a,ChandH_06a,LevshakovS_06b}, generally
yielding no variation in $\alpha_{\rm em}$. While detailed analyses of
HIRES ThAr spectra have shown no signs of strong systematic errors
\citep{MurphyM_01b,MurphyM_03a}, no similar general assessment has yet
been made for the bulk of the UVES constraints on
$\Delta\alpha/\alpha$.  Given the errors in the ESO ThAr line-list, it
is imperative to do so.

This paper is organised as follows. In Section \ref{sec:synth} the
various ThAr transition tables in the literature are combined
according to their different strengths and weaknesses. Section
\ref{sec:select} details the line selection algorithm. Section
\ref{sec:final} describes the final ThAr line-list for use in
calibrating VLT/UVES spectra and shows the distortions of the
wavelength scale introduced by the errors in the ESO line-list.
Section \ref{sec:alpha} assesses the general implications these
distortions have for current UVES constraints on a varying
fine-structure constant.

\section{Synthesis of existing Th and Ar line-lists}\label{sec:synth}

Before selecting which ThAr lines are to be used to calibrate UVES, we
first require a list of laboratory wavelengths for as many features as
possible which appear in measured UVES ThAr lamp spectra.  Most
features, especially below $7000$\,\AA, are either known to be due to
Th or are unidentified (but probably due to Th), while $\sim$7\,per
cent of the features are from Ar and $<1$\,per cent are from
`contaminant' species such as Na{\sc \,i}, Mg{\sc \,i}, Ca{\sc \,ii}
and Fe{\sc \,i}. The Th and Ar lines, and even some of the contaminant
lines, are catalogued in various atlases derived from painstaking (but
usually rather old) laboratory work. These atlases provide our
starting point. However, every ThAr lamp gives a somewhat different
spectrum and there always appear additional lines which cannot be
found in any atlas, even as `unidentified' lines. These are probably
due to additional contaminant ions and molecular species in the lamp.
This means that our prior knowledge of the ThAr spectrum from any
given lamp is, at best, incomplete, thereby necessitating the
selection procedure detailed in the next section.

There is no single ThAr atlas covering the whole wavelength range of
UVES, $\sim$3030--10540\,\AA. However, there is one atlas of Th
lines only which does cover this range, \citet[hereafter
\citetalias{PalmerB_83a}]{PalmerB_83a}. The \citetalias{PalmerB_83a}
absolute velocity precision varies from $\sim$15 to $\sim$120\,\ms\
depending on the line intensity.  \citetalias{PalmerB_83a} identify a
large number of Th{\sc \,i}, {\sc \,ii} and {\sc \,iii} lines in their
spectrum but a large number of unidentified lines remain, most of
which are probably due to Th.  Furthermore, on comparison with real
ThAr lamp spectra, one notices that there are still several thousand
lines which cannot be accounted for by Ar or contaminant species.

Very recently, \citet[hereafter \citetalias{LovisC_07a}]{LovisC_07a}
used a large library of ThAr spectra taken over one month with the
High Accuracy Radial Velocity Planet Searcher (HARPS) spectrograph on
the 3.6-m ESO telescope to improve the measurement precision of many
ThAr lines, at least over the HARPS wavelength range of
$3785$--$6915$\,\AA. They identified in their spectra lines from the
\citetalias{PalmerB_83a} atlas which showed no strong wavelength
variations with time and obtained line positions with a
root-mean-square variation (RMS) of typically $5$--$10\,\ms$.  Thus,
while bootstrapping their overall wavelength scale to that of
\citetalias{PalmerB_83a}, they were able to correct the wavelengths of
individual \citetalias{PalmerB_83a} lines, especially weaker lines.
Moreover, they were able to measure the wavelengths of all other
features in their ThAr spectra, again at the $5$--$10\,\ms$ precision
level. They removed from their list lines which saturated their CCD,
(very few) lines which appeared in the \citetalias{PalmerB_83a}
catalogue but which were not detected in the HARPS spectra and, most
importantly, they removed lines which were either closely blended with
other lines at the HARPS resolving power of $R=120000$ or were
observed to change position with time. The latter indicates either
that the line experiences significant shifts with changing lamp
pressure or current or that the line is actually a blend and that the
relative intensities of the blended lines varies with changing lamp
conditions. In this work we use all lines from the
\citetalias{LovisC_07a} atlas excluding those which have relative
wavelength errors amounting to $>120\,\ms$; a total of 8219 out of
8442 lines are used from \citetalias{LovisC_07a}.

The \citetalias{LovisC_07a} catalogue ranges from $3785$ to
$6915$\,\AA. In selecting lines for use in calibrating UVES, we use
the \citetalias{PalmerB_83a} line list for Th and contaminant lines
outside this range. However, it is important to begin with a list
which includes the known Ar lines as well so that (i) the maximum
number of known blends can be identified and removed and (ii) the
maximum number of lines are available for calibrating UVES. Three Ar
line-lists exist with similar wavelength precision -- \citet[hereafter
\citetalias{NorlenG_73a}]{NorlenG_73a}, \citet[hereafter
\citetalias{WhalingW_95a}]{WhalingW_95a}, and \citet[hereafter
\citetalias{WhalingW_02a}]{WhalingW_02a} -- each of which has
different advantages and disadvantages. The \citetalias{NorlenG_73a}
atlas contains both Ar{\sc \,i} and Ar{\sc \,ii} lines whereas
\citetalias{WhalingW_95a} contains Ar{\sc \,ii} lines only and
\citetalias{WhalingW_02a} contains Ar{\sc \,i} lines only. Fewer lines
are listed by \citetalias{NorlenG_73a} compared to
\citetalias{WhalingW_95a} and \citetalias{WhalingW_02a} since the
\citetalias{NorlenG_73a} experiment was less sensitive to weak lines.
The velocity precision achieved by \citetalias{NorlenG_73a} for the
Ar{\sc \,i} lines is better than that of \citetalias{WhalingW_02a} but
the velocity precision of the \citetalias{WhalingW_95a} Ar{\sc \,ii}
lines is better than \citetalias{NorlenG_73a}'s. Finally, there is a
calibration difference such that the \citeauthor{WhalingW_95a}
wavenumbers are larger than \citeauthor{NorlenG_73a}'s by a factor of
$S_{\rm NW} = 1 + 6.8\times10^{-8}$. \citeauthor{WhalingW_95a}'s
calibration should be more reliable since it uses calibration lines in
the same recorded spectra as the Ar{\sc \,ii} lines. Therefore, we use
the \citeauthor{WhalingW_95a} calibration scale in synthesising the Ar
line-lists.

The different lists of Ar{\sc \,i} and Ar{\sc \,ii} lines are combined
in the following ways. We do not consider the Ar{\sc \,iii} lines from
\citetalias{WhalingW_95a} since there are very few of them and since
their wavelengths have larger uncertainties (W.~Whaling, private
communication).
\begin{itemize}
\item Ar{\sc \,i}: Following the recommendation in
  \citetalias{WhalingW_02a} we used \citetalias{NorlenG_73a}
  wavenumbers (scaled by $S_{\rm NW}$) when available, otherwise
  \citetalias{WhalingW_02a} values were used. For
  \citetalias{NorlenG_73a} lines, the intensity scale of
  \citetalias{NorlenG_73a} was used as re-cast by
  \citetalias{DeCuyperJ-P_98a} onto a logarithmic scale. For
  \citetalias{WhalingW_02a} lines, the \citetalias{WhalingW_02a}
  intensity scale was used.
\item Ar{\sc \,ii}: \citetalias{WhalingW_95a} wavelengths were used
  here since \citetalias{WhalingW_95a} claim that they are relatively
  insensitive to pressure-shifts and that their values are more
  precise than \citetalias{NorlenG_73a}'s. In the few cases where
  \citetalias{NorlenG_73a} reports an Ar{\sc \,ii} line that
  \citetalias{WhalingW_95a} doesn't, the \citetalias{NorlenG_73a}
  wavelength (scaled by $S_{\rm NW}$) was used. For
  \citetalias{WhalingW_95a} lines the \citetalias{WhalingW_95a}
  intensity scale was used (this is the same as the
  \citetalias{WhalingW_02a} intensity scale). For
  \citetalias{NorlenG_73a} values, the intensity scale of
  \citetalias{NorlenG_73a} was used as re-cast by
  \citetalias{DeCuyperJ-P_98a} onto a logarithmic scale. Note that
  \citetalias{NorlenG_73a} uses the intensity scale of
  \citet{MinnhagenL_73a} for Ar{\sc \,ii} lines above air wavelengths
  of $\lambda_{\rm air}=7600$\,\AA\ and below $\lambda_{\rm
    air}=3400$\,\AA.  This subtlety is neglected in the considerations
  below.
\end{itemize}

After combining the \citetalias{LovisC_07a}, \citetalias{PalmerB_83a},
\citetalias{NorlenG_73a}, \citetalias{WhalingW_95a} and
\citetalias{WhalingW_02a} lists in the above way we obtain the list
hereafter referred to as LPWN. Note from the above that the intensity
scales for each species (e.g. Th{\sc \,i}, Ar{\sc \,i}, Ar{\sc \,ii})
are different and that the relative intensities of lines from
different species depend on many factors such as lamp pressure and
current and the relative partial pressures of Th and Ar gas used.

\section{ThAr line selection}\label{sec:select}

One should not simply use the above line-list, LPWN, to calibrate UVES
spectra because of the potential for line-blending. The Fourier
transform spectra of \citetalias{PalmerB_83a},
\citetalias{NorlenG_73a}, \citetalias{WhalingW_95a} and
\citetalias{WhalingW_02a} and the HARPS spectra of
\citetalias{LovisC_07a} all have resolving powers $R \ge 120000$
whereas the UVES resolving power is typically $35000$--$70000$ and
possibly as high as $110000$. Therefore, a major part of the line
selection that follows is the rejection of `close' blends. The concept
of `close' clearly depends on the relative intensities of the blended
lines and so it is necessary to put all lines in LPWN list onto the
one intensity scale typical of UVES ThAr spectra.  Furthermore, since
there are always additional unknown lines in measured ThAr spectra, a
given ThAr line can only be used for calibration if it is {\it
  measured} to be reliable in the UVES ThAr spectrum. For these
reasons we have constructed a UVES ThAr spectrum for use in the line
selection process.

\subsection{UVES ThAr spectrum}\label{ssec:uves_thar}

ThAr spectra taken as part of regular UVES maintenance activities with
a $0\farcs6$ slit and no CCD rebinning were retrieved from the ESO
Science Archive. Exposures from several different standard UVES
wavelength settings\footnote{The 760-nm exposure was taken on the 21st
  March 2005. All other exposures were taken on the 11th February
  2004.}  (346, 437, 580, 600, 760 and 860-nm) were used to provide
complete wavelength coverage, with the exception of three small
echelle order gaps redwards of $10000$\,\AA\
($10084.43$--$10085.93$\,\AA, $10252.42$--$10256.95$\,\AA\ and
$10426.10$--$10433.76$\,\AA). The UVES
pipeline\footnote{http://www.eso.org/projects/esomidas} was used to
extract and wavelength calibrate the data.  Several modifications to
the pipeline, which will be reported elsewhere, were made to improve
object extraction and wavelength calibration. The data were
re-dispersed to a log-linear wavelength scale using {\sc uves
  popler}\footnote{http://www.ast.cam.ac.uk/$\sim$mim/UVES\_popler.html},
code specifically written to re-disperse and combine UVES data from
multiple wavelength settings. Overlapping regions of spectra were cut
away so that only one raw exposure contributed to the final spectrum
at any given wavelength. No effort was made to flux-calibrate the
final spectrum and the blaze-function of the echelle grating was still
evident in the combined spectrum. However, our results from the
intensity re-scaling in the next section demonstrate that this is a
minor consideration. The final spectrum has a resolving power of
$R\approx70000$ and the log-linear dispersion is set to $1.75\,\kms$.

\subsection{ThAr line selection algorithm}\label{ssec:algor}

The above line-list, LPWN, is treated as the input catalogue to the
following algorithm for selecting a final list for calibration of
UVES. We make three passes through the algorithm, refining at each
pass the line list used to calibrate the UVES ThAr spectrum and other
parameters specified below. Figures \ref{fig:alpha}--\ref{fig:comp}
pertain to the final pass through the algorithm.

\begin{enumerate}
  \begin{figure*}
  \centerline{
   \hbox{
    \includegraphics[height=0.4\textwidth,angle=270]{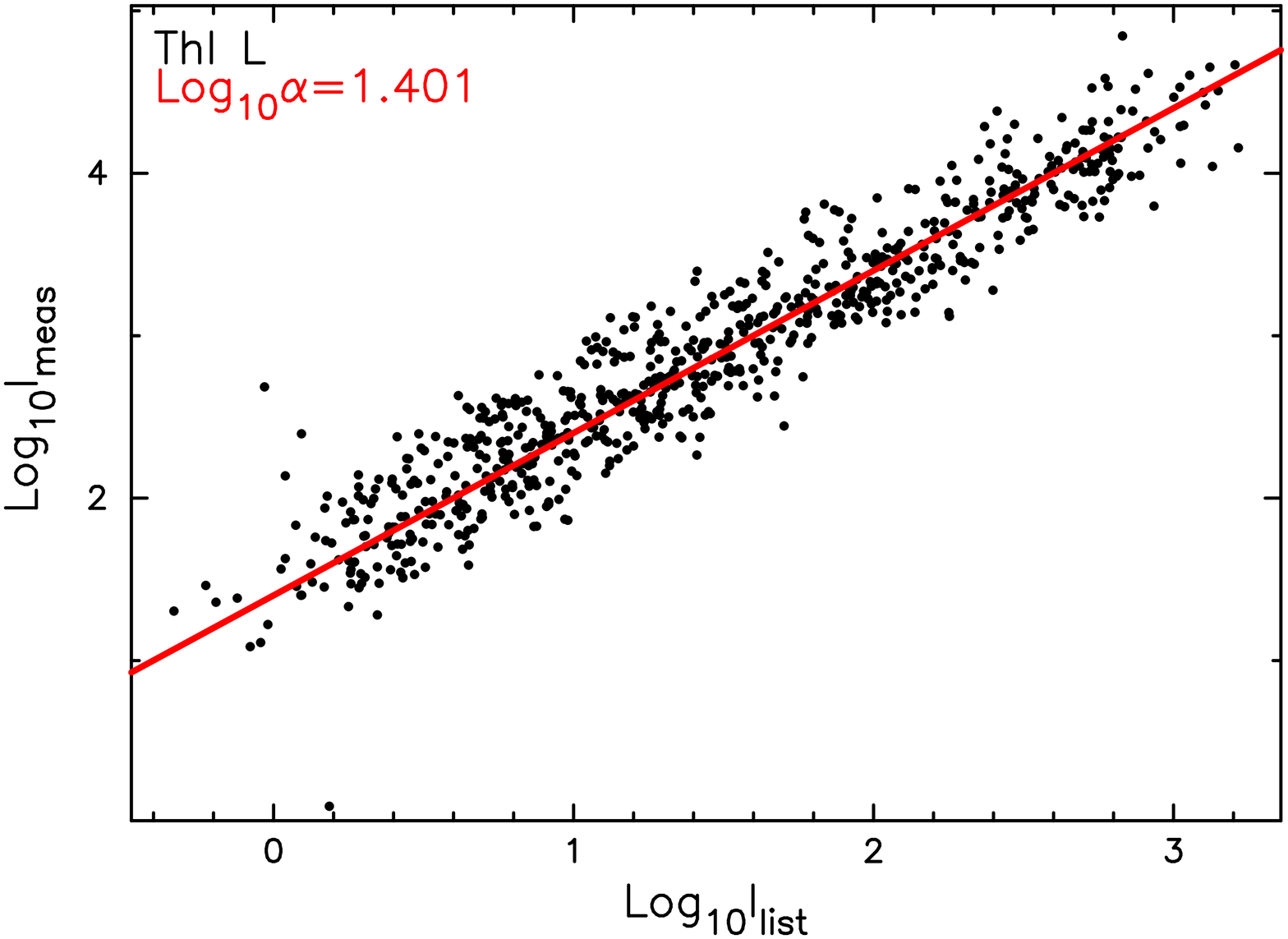}
    \hspace{0.01\textwidth}
    \includegraphics[height=0.4\textwidth,angle=270]{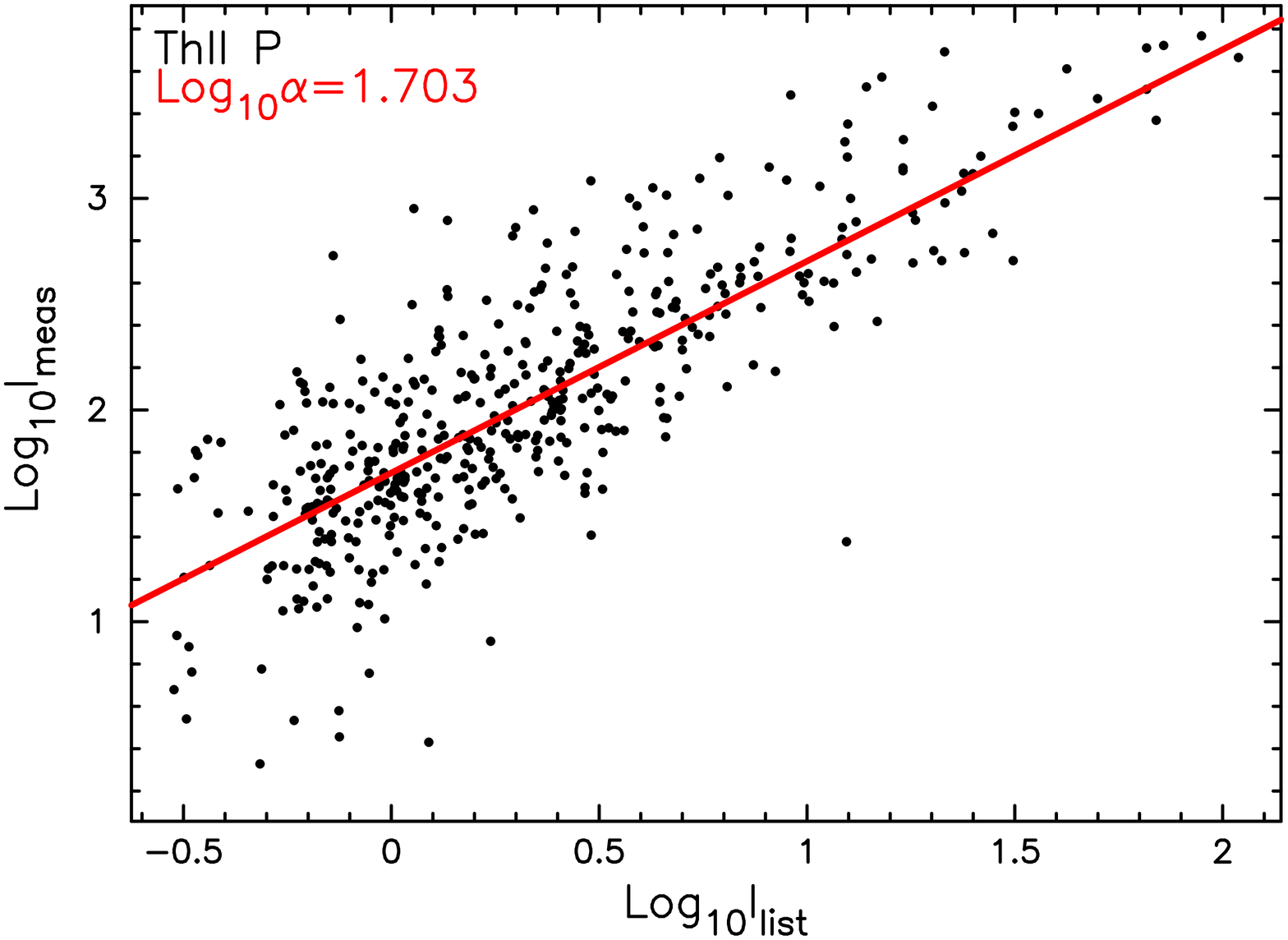}
   }
  }
  \caption{Examples of the intensity scaling procedure. Measured and
    catalogued intensities, $I_{\rm meas}$ and $I_{\rm list}$
    respectively, for unblended lines from each category -- Th{\sc
      \,i} from \citetalias{LovisC_07a} and Th{\sc \,ii} from
    \citetalias{PalmerB_83a} in this figure -- are compared to derive
    the scale-factor, $\alpha$. All lines from that category are then
    scaled by $\alpha$ to the UVES intensity scale.}
  \label{fig:alpha}
  \end{figure*}

\item Wavelength calibrating the UVES ThAr spectrum: The procedure in
  Section \ref{ssec:uves_thar} for constructing the UVES ThAr spectrum
  requires an input line-list for the wavelength calibration stage. At
  the first pass we use the line-list provided by ESO with the UVES
  pipeline. At the second and third passes the output from the
  previous pass through the algorithm is used. Note that the line-list
  used at the first pass is irrelevant; only an approximate wavelength
  scale needs to be established initially and this is greatly refined
  by subsequent passes through the algorithm.

\item Gaussian fitting: Each line in the input list LPWN is searched
  for in the UVES ThAr spectrum and fitted with a Gaussian. The fit
  includes the 13 pixels centred on the pixel with maximum intensity.
  A first-guess continuum level is defined by averaging the first and
  last 2 pixels in the window. A first-guess continuum slope is
  defined from the difference between these averages. A first-guess
  amplitude is defined as the maximum intensity minus the continuum
  level. A first-guess width is defined from the velocity difference
  between the first and last pixels which have intensities less than
  the continuum plus half the first-guess amplitude. The first-guess
  central velocity is calculated as the intensity-weighted velocity of
  the 3 central pixels. A 5-parameter Gaussian fit is performed and
  the best-fit parameters are used in the subsequent steps of the line
  selection algorithm. The initial guesses and Gaussian fitting
  procedure are identical to those we used in our modifications to the
  UVES pipeline.

\item Intensity re-scaling: The different Th and Ar atlases combined
  above have different intensity scales. Furthermore, one would not
  expect to find the same relative intensities for lines of different
  ionization stages (e.g.~Ar{\sc \,i} and Ar{\sc \,ii}) in the
  laboratory and UVES spectra since the ThAr lamps will have had
  different operating conditions (e.g.~pressure, current etc.). We
  therefore aim to place all lines from the different ionic species
  (i.e.~Th{\sc \,i}/{\sc ii}/{\sc iii}, Ar{\sc \,i}/{\sc ii},
  contaminant and unidentified species) from each different intensity
  scale (i.e.~\citetalias{LovisC_07a}, \citetalias{PalmerB_83a},
  \citetalias{NorlenG_73a} and
  \citetalias{WhalingW_95a}/\citetalias{WhalingW_02a}) on a single
  intensity scale directly related to the UVES ThAr spectrum. First,
  any pairs of lines within 13 km/s of each other are removed from the
  input line list LPWN. For each `category' of line -- for example,
  Th{\sc \,i} from \citetalias{LovisC_07a}, Th{\sc \,i} from
  \citetalias{PalmerB_83a}, Ar{\sc \,ii} from \citetalias{LovisC_07a},
  Ar{\sc \,ii} from \citetalias{WhalingW_95a} etc. -- the catalogued
  intensities are compared with those measured from the Gaussian fits
  performed on the UVES ThAr spectrum. The median ratio of measured
  and catalogued intensity, $\alpha$, is defined as the scale-factor.
  All lines from this category are then scaled to the UVES intensity
  scale by multiplying their catalogued intensities by $\alpha$. Two
  examples of this process are shown in Fig.~\ref{fig:alpha}. Figure
  \ref{fig:inten} shows all lines from all categories placed on the
  UVES intensity scale.

  \begin{figure}
  \centerline{\includegraphics[height=0.95\columnwidth,angle=270]{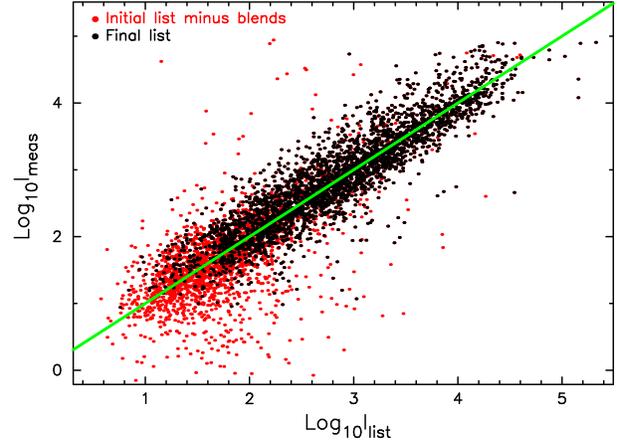}}
  \caption{Lines from all categories placed on the UVES intensity
    scale. The grey/red points are lines from the initial LPWN list
    which satisfy the blending criterion defined in step (iv) of the
    ThAr selection algorithm. The black points are lines satisfying
    all selection criteria and constitute the final ThAr line-list.
    Here, $I_{\rm list}$ refers to the re-scaled listed intensities
    and $I_{\rm meas}$ is the amplitude of the line measured from the
    UVES ThAr spectrum.}
  \label{fig:inten}
  \end{figure}

  \begin{figure*}
  \centerline{
   \hbox{
    \includegraphics[width=0.38\textwidth,angle=270]{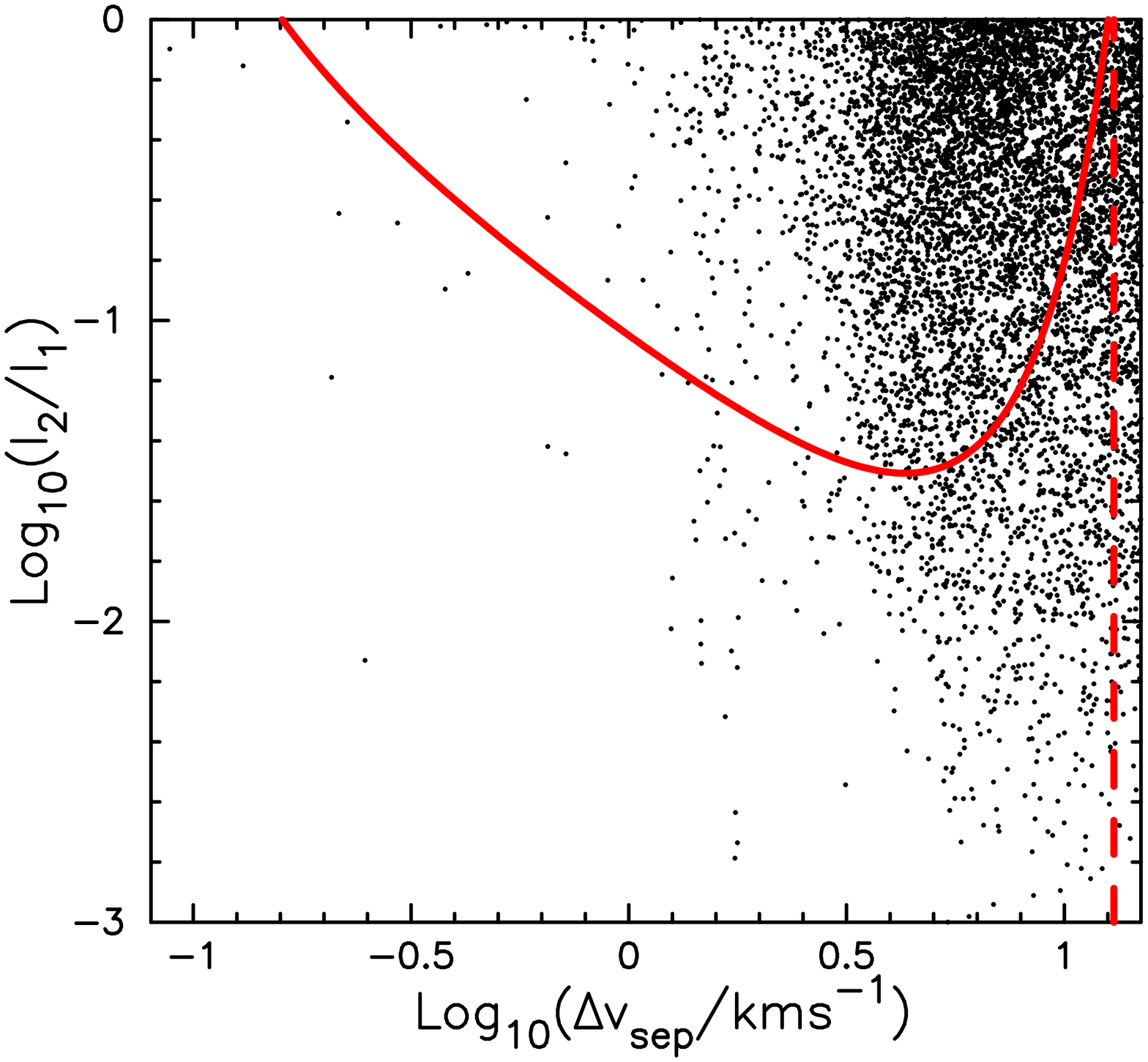}
    \hspace{0.01\textwidth}
    \includegraphics[width=0.38\textwidth,angle=270]{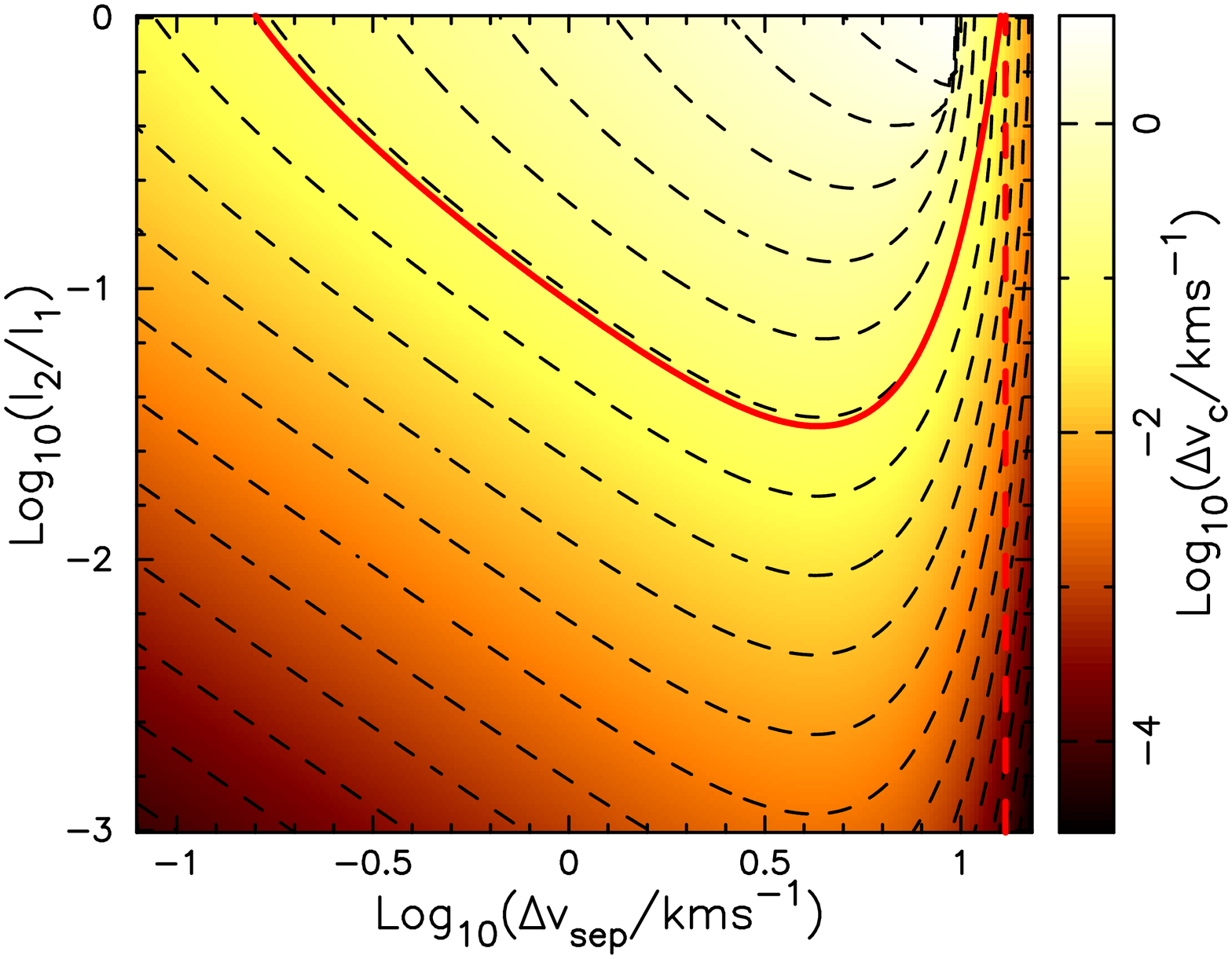}
   }
  }
  \caption{Removal of strongly blended lines. {\it Right}: The shift
    in the centroid of a synthetic Gaussian due to blending with a
    weaker line. The black dashed lines are contours equally spaced in
    the logarithm of the shift. The solid grey/red line is a simple
    by-eye fit to these contours at a velocity tolerance of $40\,\ms$
    and $R=43000$ [equation (\ref{eq:blend})].  All lines with
    blending lines more than 13 km/s away are safe from velocity
    shifts greater than $40\,\ms$ and this is marked with the vertical
    dashed grey/red line. This figure is available in colour in the
    online version of the journal on {\it Synergy} and from
    http://www.ast.cam.ac.uk/$\sim$mim/pub.html. {\it Left}: Pairs of
    lines in the input line-list LPWN with the same solid and dashed
    grey/red lines from the right-hand panel.}
  \label{fig:blend}
  \end{figure*}

\item Blend removal: If, when Gaussian-fitting a single line, another
  weaker blending line is present but ignored in the fit, then the
  centroid returned from the fit will be shifted towards the blending
  line. The magnitude of the shift will depend not only on the
  velocity separation between the two lines, $\Delta v_{\rm sep}$, but
  also on the relative intensities of the two lines, $I_2/I_1$. When
  the two lines are not resolved from each other, it is easy to
  estimate the velocity shift: the new centroid wavelength is
  approximately the intensity-weighted mean wavelength of the two
  blended lines,
  \begin{equation}
  \lambda_{\rm c} \approx \frac{I_1\lambda_1  + I_2\lambda_2}{I_1 + I_2}\,.
  \end{equation}
  The velocity shift due to the blending line is therefore
  \begin{equation}
  \Delta v_{\rm c} \equiv \frac{\lambda_{\rm c} - \lambda_1}{\lambda_1}
  \approx \Delta v_{\rm sep}\frac{I_2/I_1}{1+I_2/I_1}\,.
  \end{equation}
  However, if the two lines are further apart and are partially
  resolved it is not clear how $\Delta v_{\rm c}$ will depend on
  $\Delta v_{\rm sep}$ and $I_2/I_1$.  Figure \ref{fig:blend}(right)
  shows the results of a numerical experiment where two blended lines
  are varied in relative intensity and separation and fitted with a
  single Gaussian with similar initial guesses as in step (ii). A
  by-eye fit to the contours of constant velocity shift gives the
  following relationship:
  \begin{equation}\label{eq:blend}
  \frac{I_2}{I_1} \approx \left(\frac{\Delta v_{\rm sep}}{\Delta
  v_{\rm tol}}-1\right)^{-1} + 0.1\left(\frac{\Delta v_{\rm sep}}{\sigma_{\rm v}}\right)^2
  + 0.0012\left(\frac{\Delta v_{\rm sep}}{\sigma_{\rm v}}\right)^4\,. 
  \end{equation}
  The first term on the right-hand-side is just the previous equation
  and applies when $\Delta v_{\rm sep}$ is small with respect to
  $\sigma_{\rm v} = (c/R)/2.355$ ($c$ is the speed of light and $c/R$
  is the spectrograph's FWHM resolution).  We therefore reject lines
  from the input line-list LPWN which have blending lines within
  $13\,\kms$ with relative intensities greater than that predicted by
  equation (\ref{eq:blend}) to produce shifts greater than a tolerance
  of $40\,\ms$ at $R=43000$. This tolerance is similar to the overall
  wavelength calibration residuals achievable at a typical UVES
  resolution of $R=43000$ with the final ThAr list.

\item Removal of weak lines: If ThAr lines appear weak in the UVES
  lamp spectrum then the UVES pipeline's line-identification algorithm
  can fail or yield a false identification. The velocity precision
  available in very weak lines is also too small to be useful in
  calibrating the UVES wavelength scale. We therefore reject any lines
  with measured intensities (above the measured continuum) less than 4
  times the measured continuum level.

\item FWHM selection: Any additional unknown features in the UVES ThAr
  spectrum can cause blending with the lines remaining after steps
  (iv) \& (v) above. The next three steps aim to reject those lines
  which are effected in this way. The first of these steps is to
  remove lines whose widths are clearly inconsistent with the
  instrumental resolution, in this case $R\approx70000$ or ${\rm
    FWHM}\approx4.3\,\kms$.  Figure \ref{fig:fwhm} shows the
  distribution of ${\rm FWHM}$ for all lines which survive steps (iv)
  \& (v) above. After visual inspection of the lines lying away from
  the main cluster around ${\rm FWHM}\!\sim\!4.3\,\kms$ it was clear that
  lines wider than $\sim5.3$\,\kms\ or narrower than $\sim$3.5\,\kms\
  should be removed. Lines were too wide when they were blended with
  other unknown features or where saturation of the CCD occurred.
  Lines were typically fitted as very narrow when they were very weak.

  \begin{figure}
  \centerline{\includegraphics[height=0.95\columnwidth,angle=270]{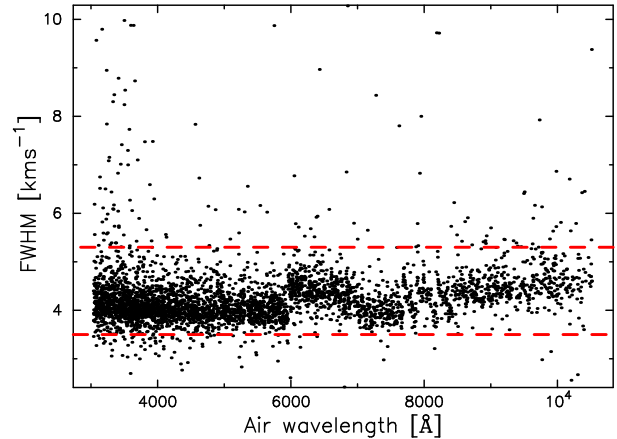}}
  \caption{The FWHM distribution of the lines surviving steps (iv) \&
    (v) in the selection algorithm. Most lines lie in a well-defined
    band around the expected ${\rm FWHM}$ of $\sim$4.3\,\ms. The
    dashed lines indicate our selection criteria for suitable
    calibration lines. The structure seen in this band is due to
    slightly differing resolutions in the different exposures from
    different wavelength settings.}
  \label{fig:fwhm}
  \end{figure}

\item High slope rejection: One of the improvements we made to the
  UVES pipeline was the addition of a continuum slope parameter to the
  Gaussian fitting of ThAr lines during wavelength calibration. If a
  line is close to a strong, previously unknown feature in the UVES
  ThAr spectrum then a large slope (relative to the line's intensity)
  might be needed to fit the line properly. Since these unknown
  features probably vary with lamp conditions, lines fitted with large
  continuum slopes are best avoided when calibrating the wavelength
  scale. Figure \ref{fig:slope} shows the distribution of the absolute
  velocity difference, $\Delta v_{\rm s}$, between two Gaussian fits
  -- one made including the slope parameter, the other made with the
  slope fixed to zero -- with continuum slope normalized by the line
  intensity. Visual inspection of lines with $\Delta v_{\rm s}$
  greater than $80\,\ms$ showed that some of them have large
  asymmetries which are due to close blends with unknown features.
  Therefore, all lines with $\Delta v_{\rm s} \ge 80\,\ms$ are removed
  from the list at this stage.

  \begin{figure}
  \centerline{\includegraphics[height=0.95\columnwidth,angle=270]{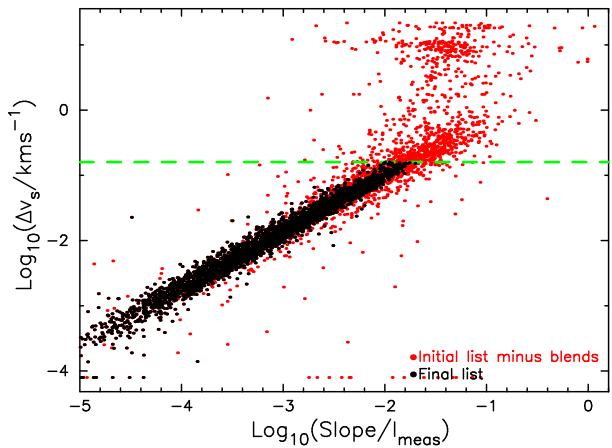}}
  \caption{Distribution of the absolute velocity difference between
    the centroids of two Gaussian fits -- one made including the slope
    parameter, the other made with the slope fixed to zero -- versus
    the normalized continuum slope. A conservative cut is made at
    $\Delta v_{\rm s}=80\,\ms$ to reject lines which might be effected
    by additional unknown features in the UVES ThAr spectrum.}
  \label{fig:slope}
  \end{figure}

\item Large residual rejection: As a final selection step we reject
  lines which are fitted at wavelengths at some variance to those
  expected from the input line list LPWN. Since the ThAr line-list
  supplied with the UVES pipeline is known to contain several
  inaccuracies (see Section \ref{sec:intro}) we do not apply this
  criterion in the first pass through the selection algorithm. In the
  second pass we reject lines which are fitted at positions more than
  $0.25\,\kms$ away from the expected position. In the third pass we
  reduce this parameter to $0.15\,\kms$, less than 1/10th of a pixel.

\end{enumerate}

\section{Final ThAr line-list}\label{sec:final}

\begin{table}
\begin{center}
  \caption{Excerpt from the final ThAr line-list for UVES. The
    complete table is available at
    http://www.ast.cam.ac.uk/$\sim$mim/pub.html and in the online
    version of this paper. The first column is the wavenumber
    ($\lambda_{\rm vac} = 1\times10^8/\omega_0$), the second column is
    the air-wavelength computed from the vacuum wavelength,
    $\lambda_{\rm vac}$, using the \citet{EdlenB_66a} formula for the
    dispersion of air at $15^\circ$C and atmospheric pressure
    \citep[see][for detailed discussion]{MurphyM_01b}. The third
    column is the logarithm of the original listed intensity
    normalized to the (arbitrary) intensity scale of the UVES ThAr
    spectrum. The fourth and fifth columns provide the line
    identification when available. The final column identifies the
    source of the wavenumber and which intensity scale the line was
    originally on: `L' indicates \citetalias{LovisC_07a}, `P'
    indicates \citetalias{PalmerB_83a}, `N' indicates
    \citetalias{NorlenG_73a} while `W' indicates
    \citetalias{WhalingW_95a} or \citetalias{WhalingW_02a}.
    Unidentified lines are generally given element designations `XX'
    and ionization levels of `0'. For unidentified lines in
    \citetalias{LovisC_07a} an ionization level `0' is given if the
    line appears in PE83 or `1' if \citetalias{LovisC_07a} claim the
    line was previously unknown.}
\label{tab:list}
\vspace{-2mm}
\begin{tabular}{cccccc}\hline
$\omega_0$        & $\lambda_{\rm air}$ & $\log_{10}I$ & \multicolumn{2}{c}{Line Identity} & Source  \\
$[{\rm cm}^{-1}]$ & $[$\AA$]$           &              & Element        & Ion              &         \\\hline
26473.6290        & 3776.271162 & 3.299 & Th & {\sc i}   & P \\
26450.5761        & 3779.562436 & 1.634 & XX & 0         & P \\
26426.4527        & 3783.012697 & 2.576 & Th & {\sc ii}  & P \\
26424.4709        & 3783.296425 & 2.935 & Th & {\sc ii}  & P \\
26415.5397        & 3784.575599 & 2.242 & XX & 0         & L \\
26408.3869        & 3785.600691 & 2.644 & Th & {\sc ii}  & L \\
26402.9308        & 3786.383000 & 2.653 & Ar & {\sc ii}  & L \\
26392.1220        & 3787.933732 & 1.813 & Th & {\sc ii}  & L \\\hline
\end{tabular}
\end{center}
\end{table}

\begin{figure*}
\centerline{
 \hbox{
  \includegraphics[height=0.40\textwidth,angle=270]{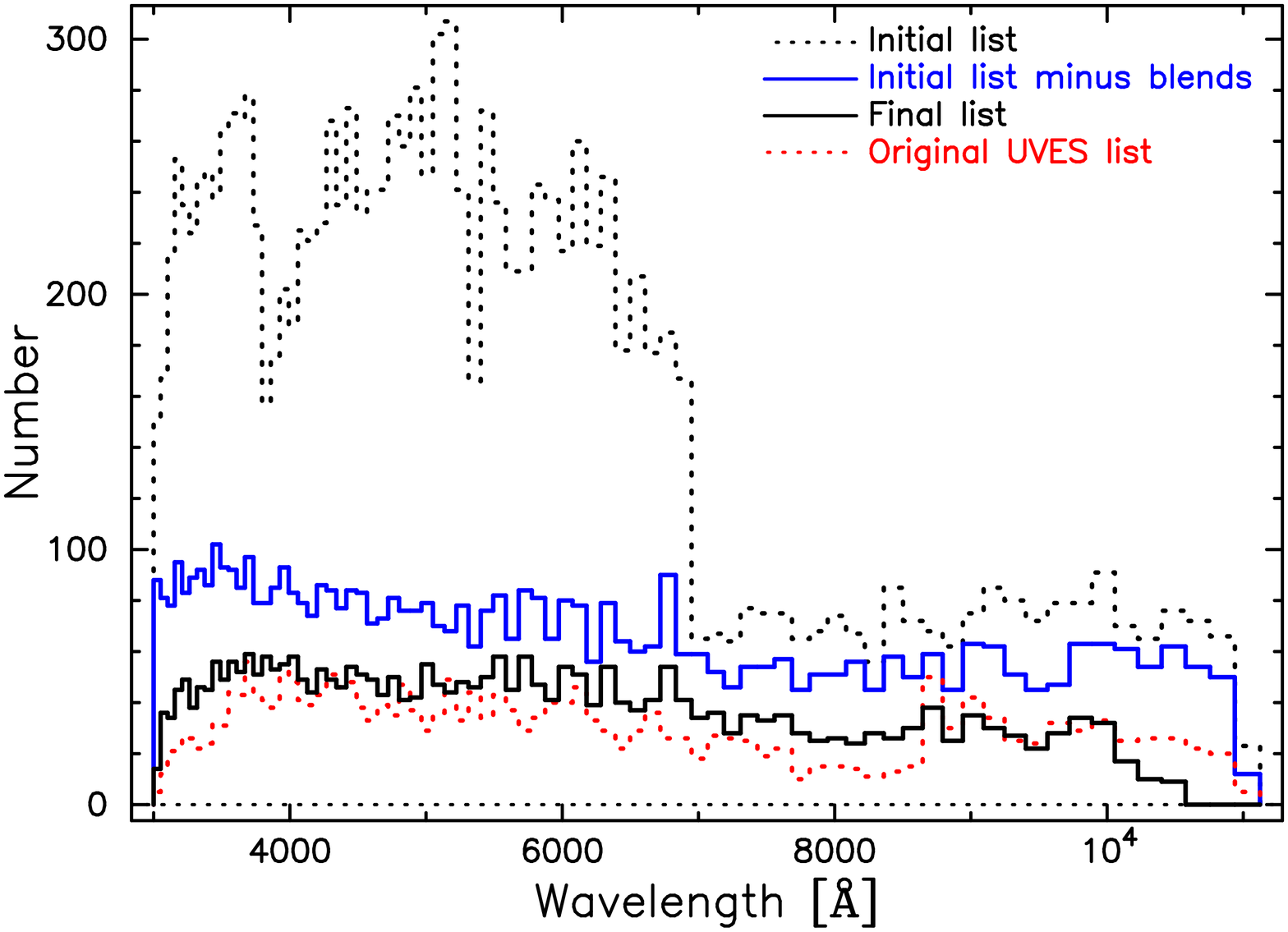}
  \hspace{0.01\textwidth}
  \includegraphics[height=0.40\textwidth,angle=270]{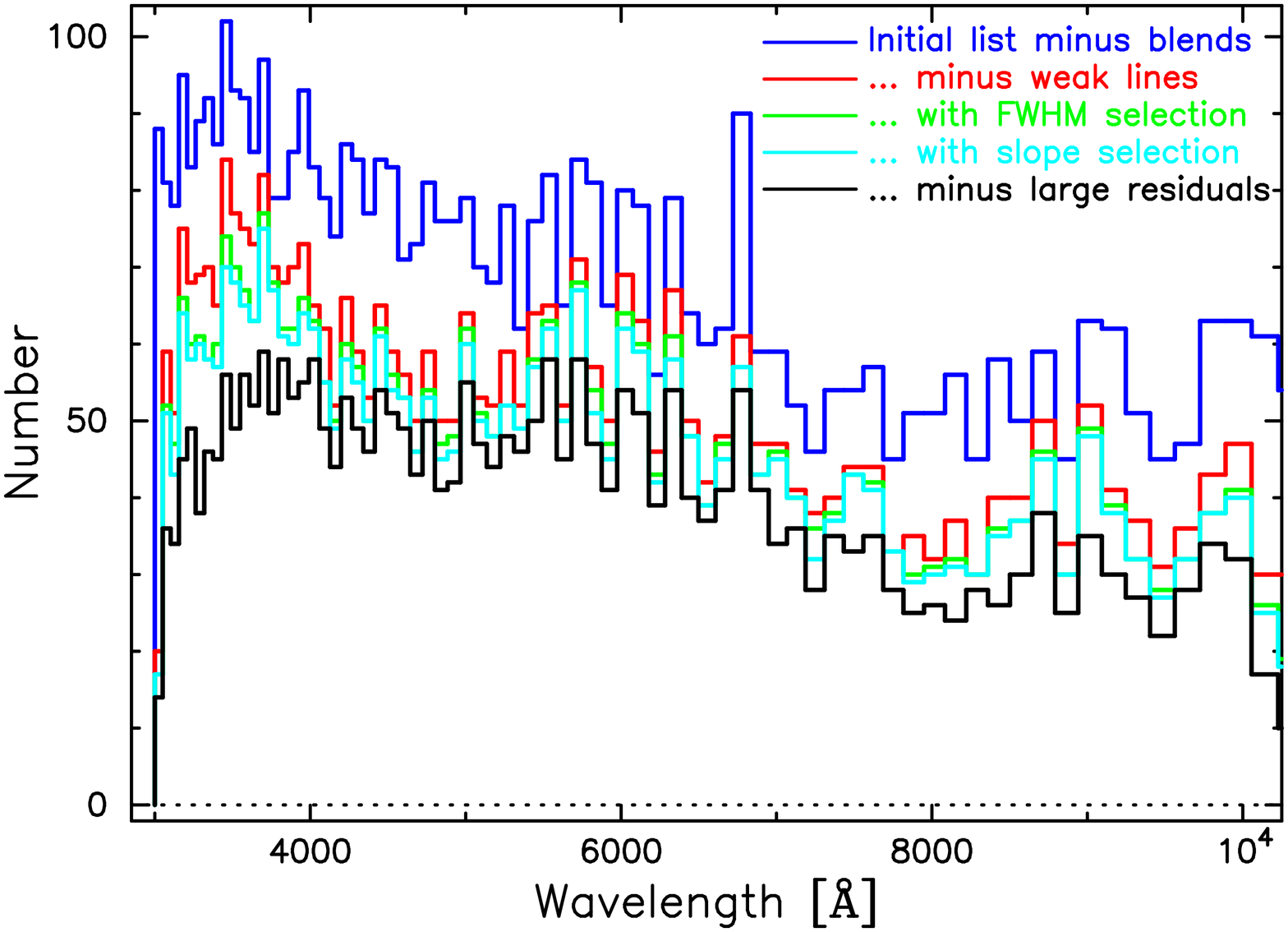}
 }
}
\caption{Histograms showing the expected number of lines per echelle
  order for different line-lists. {\it Left}: Comparison of the
  initial line-list with the list after close blends have been
  removed. Also shown is a comparison of the final line-list with that
  provided with the UVES pipeline. {\it Right}: The reduction of the
  line-list through the various selection stages.}
\label{fig:select}
\end{figure*}

The ThAr line-list formed after the third pass through the selection
algorithm, referred to hereafter as the final or new list, is shown in
part in Table \ref{tab:list} and is available in its entirety from
http://www.ast.cam.ac.uk/$\sim$mim/pub.html and in the electronic
version of this paper. In addition to an ASCII version, we also
provide a version in a format which is readable by the UVES pipeline
distributed by ESO.

The initial input list LPWN contains 13903 lines while the final list
contains just 3070. For comparison, the line-list supplied with the
UVES pipeline contains 2387 lines over the wavelength range covered by
UVES. So, although a large number of lines are rejected in the
selection algorithm, there are certainly enough remaining lines to
provide a reliable calibration of the UVES wavelength scale. Figure
\ref{fig:select}(left) shows the distribution of lines with wavelength
in bins which approximate the size of extracted UVES echelle orders.
Compared to the initial input list, the distribution of lines is quite
uniform. This is mainly because of the rejection of close blends and
demonstrates why the LPWN catalogue should not be used without first
selecting unblended and otherwise reliable lines. Of course, since
LPWN contains so many lines, one can always achieve very low {\it
  formal} wavelength calibration residuals simply by defining a
sufficiently small rejection threshold during the calibration.
However, it is important to realise that with this approach many of
the lines eventually used for calibration will actually be blends
which will distort the wavelength solution.

\begin{figure*}
\centerline{
 \hbox{
  \includegraphics[height=0.40\textwidth,angle=270]{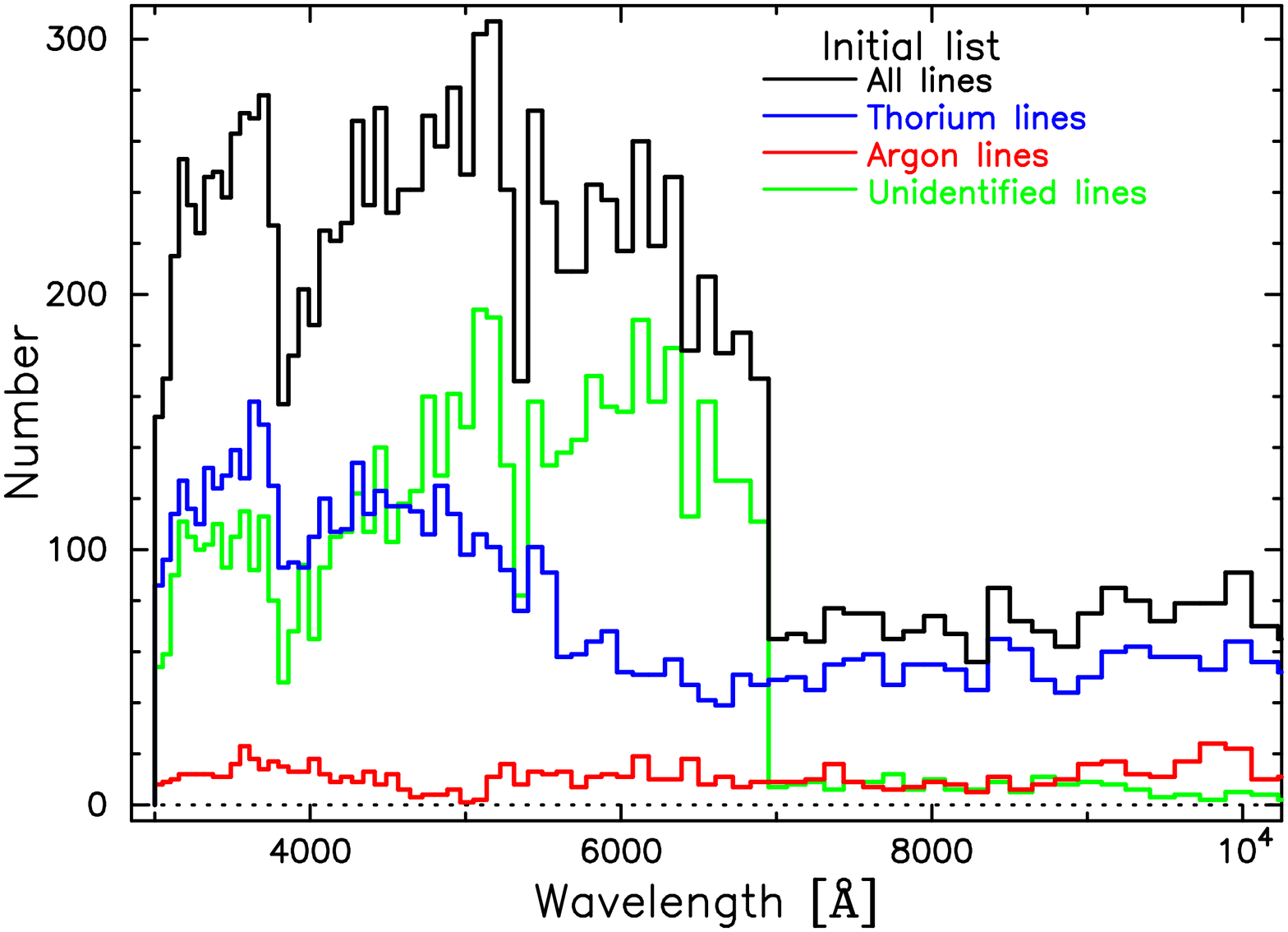}
  \hspace{0.01\textwidth}
  \includegraphics[height=0.40\textwidth,angle=270]{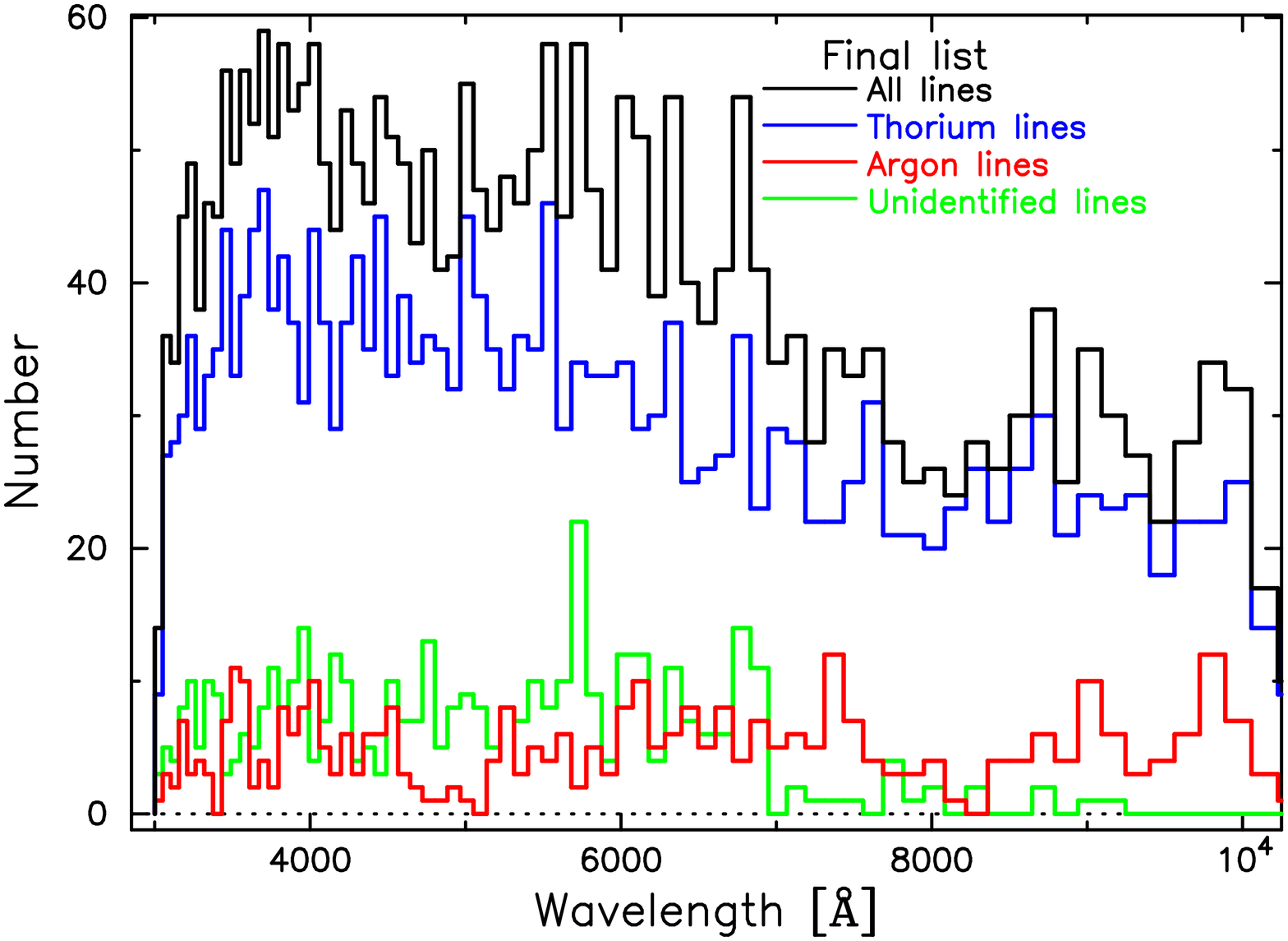}
 }
}
\caption{Histograms showing the contributions from Th, Ar and
  unidentified lines to the initial (left panel) and final (right
  panel) ThAr line-lists.}
\label{fig:comp}
\end{figure*}

Figure \ref{fig:select}(left) also shows that there are always more
than $\sim$20 useful lines per UVES echelle order in the final list.
This is enough to supply a robust wavelength calibration solution.
Figure \ref{fig:select}(right) shows the distribution of lines at each
step in the third pass through the selection algorithm. After the
removal of close blends none of the subsequent steps removes lines in
a strongly wavelength-dependent manner, as expected. Figure
\ref{fig:comp}(left) shows the contribution from Th, Ar and
unidentified lines in the initial input list LPWN while
Fig.~\ref{fig:comp}(right) shows the composition of the final list.
Note the strong decrease in the fraction of unidentified lines through
the selection algorithm. This is because most of the unidentified
lines are those newly discovered by \citetalias{LovisC_07a}, many of
which are quite weak and/or blended at the UVES resolution.

\begin{figure}
\centerline{\includegraphics[width=0.95\columnwidth]{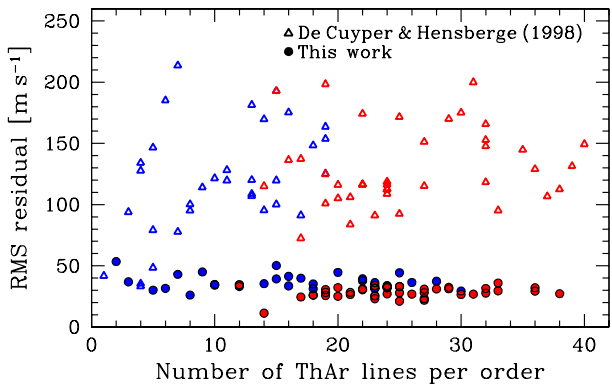}}
\caption{Improvement of the wavelength calibration solution using the
  new line list (closed circles) over that achievable with the
  $R=50000$ list from \citetalias{DeCuyperJ-P_98a} (open triangles).
  The blue/darker points represent echelle orders from the 346-nm
  standard UVES wavelength setting while the red/lighter points are
  from the 580-nm setting. The calibration residuals for the 580-nm
  setting using our new list are slightly better than for the 346-nm
  setting due to the improved precision in the new
  \citetalias{LovisC_07a} ThAr atlas.}
\label{fig:wavres}
\end{figure}

We have used the final list to calibrate the UVES ThAr spectrum a
final time and the resulting spectrum is available at
http://www.ast.cam.ac.uk/$\sim$mim/pub.html and in the electronic
version of this paper. In extracting and wavelength calibrating the
spectrum with the modified UVES pipeline, we noted a dramatic
improvement in the wavelength calibration residuals. To quantify this
we wavelength calibrated the ThAr spectrum with two different input
line-lists -- the final list derived above and the selection of
\citetalias{DeCuyperJ-P_98a} for $R=50000$ -- with the same pipeline
parameters. The one exception was the tolerance level for rejecting
discrepant lines in the pipeline's wavelength calibration step. This
was adjusted to optimize the calibration for each list; i.e.~a balance
was sought between the number of calibration lines and RMS residual
per echelle order. The results are summarized in Fig.~\ref{fig:wavres}
which compares these two quantities for each echelle order in the 346-
and 580-nm standard UVES wavelength settings.

Fig.~\ref{fig:wavres} demonstrates that our new list provides a factor
of $\ga$3 improvement in the wavelength calibration residuals and,
particularly for the 346-nm setting, provides more lines per echelle
order on average. Note also the significant reduction in the number of
orders -- all at the bluest extreme of the 346-nm setting -- which
have less than 10 lines for calibration when using the new list. The
346- and 580-nm settings also illustrate the impact of the new
\citetalias{LovisC_07a} ThAr catalogue on the residuals: the 346-nm
setting covers wavelengths not covered by HARPS (spectra from which
the \citetalias{LovisC_07a} catalogue was derived) while the 580-nm
setting uses lines almost exclusively from \citetalias{LovisC_07a}.
The mean residual for the 346-nm setting is 36\,\ms\ while that for
the 580 setting is 28\,\ms\ -- a small but clear improvement due to
the improved line wavelengths in \citetalias{LovisC_07a}. Note that
the mean residuals in the two settings are the same (130\,\ms) when
the spectra are calibrated with the \citetalias{DeCuyperJ-P_98a} list.

\begin{figure}
\centerline{\includegraphics[height=0.95\columnwidth,angle=270]{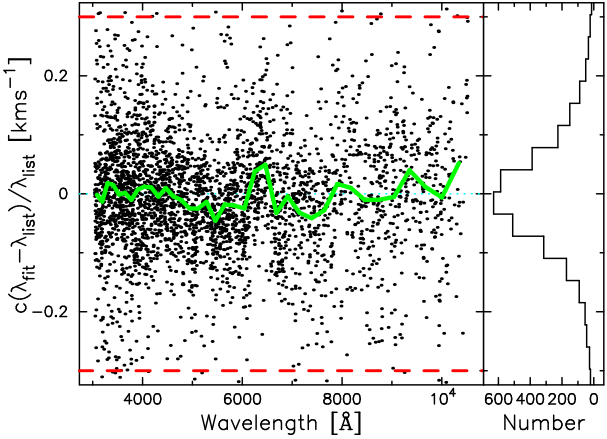}}
\caption{Residuals between the Gaussian fit centroid wavelengths
  measured from the ThAr spectrum (reduced using the unmodified UVES
  pipeline and ESO line-list), $\lambda_{\rm fit}$, and the
  wavelengths listed in the LPWN line-list compiled in this work,
  $\lambda_{\rm list}$. Note the structure in the residuals as a
  function of wavelength. The mean residual, in bins which approximate
  the size of two UVES echelle orders, is shown as a grey/green thick
  line. The typical error in the mean residual in each bin is
  $\sim$5--10\,\ms. Thus, the strong variations from bin to bin and
  over longer wavelength ranges are statistically significant and
  indicate distortions of the wavelength scale introduced by the
  various inaccuracies in the ESO line-list and the UVES pipeline
  wavelength calibration procedures.}
\label{fig:ESO}
\end{figure}

Part of the motivation for this work was the various inaccuracies we
noticed in the ThAr line-list distributed by ESO with the UVES
pipeline (see Section \ref{sec:intro}). To investigate the effect
these inaccuracies have on reduced spectra we have reduced the same
raw ThAr exposures used above with the unmodified UVES pipeline and
the ESO line-list. We then follow the line selection algorithm in
Section \ref{ssec:algor} using a tolerance on the residuals of
$0.3\,\kms$ in step (viii). Using the LPWN line-list as input allows
us to trace any distortions of the wavelength scale introduced by the
inaccuracies in the ESO line-list and wavelength calibration
procedure. The residuals between the Gaussian fit centroid wavelengths
and the wavelengths listed in our final ThAr list derived above are
plotted in Fig.~\ref{fig:ESO}. The mean residual shows strong,
statistically significant variations with wavelength. The two most
striking features are (i) the spike in the mean residual around
6300\,\AA\ which has a peak-to-peak amplitude of $\sim$75\,\ms\ and
(ii) the large-scale decrease in the mean residual from $\sim$0\,\ms\
below $4500$\,\AA\ to $\sim$$-30\,\ms$ over the range
$5500$--$7000$\,\AA\ and then the slow increase to $\sim$$+30\,\ms$
redwards of $7000$\,\AA. These distortions of the wavelength scale
have important consequences for several applications which have used
pipeline-reduced UVES spectra. We illustrate one such application
below.

\section{Implications for current UVES constraints on a varying
  fine-structure constant}\label{sec:alpha}

Distortions of the wavelength scale of the magnitude and over the
wavelength ranges just discussed are critically important to current
UVES constraints on cosmological variations in the fine-structure
constant, $\alpha_{\rm em}$, which employ the UVES pipeline and ESO
line-list. The velocity shift, $v$, of a transition due to a small
relative variation in $\alpha_{\rm em}$, $\Delta\alpha/\alpha\ll 1$,
is determined by the $q$-coefficient for that transition:
\begin{equation}\label{eq:da}
\omega_z \equiv \omega_0 + q\left[\left(\alpha_z/\alpha_0\right)^2-1\right]\,\hspace{1em}\Rightarrow\hspace{1em}\frac{v}{c} \approx -2\frac{\Delta\alpha}{\alpha}\frac{q}{\omega_0}\,,
\end{equation}
where $\omega_0$ and $\omega_z$ are the rest-frequencies in the
laboratory and at redshift $z$, $\alpha_0$ is the laboratory value of
$\alpha_{\rm em}$ and $\alpha_z$ is the shifted value measured from an
absorber at redshift $z$. The MM method is the comparison of measured
velocity shifts from several transitions (with different
$q$-coefficients) to compute the best-fitting $\Delta\alpha/\alpha$.
Thus, the variation with wavelength in the magnitude and sign of the
velocity shift in Fig.~\ref{fig:ESO} will lead directly to spurious
measurements of $\Delta\alpha/\alpha$ depending on (i) which
transitions are utilised to compute $\Delta\alpha/\alpha$ and (ii) the
redshift of the absorption system.

An example of this problem arises in the results of \citet{ChandH_04a}
who studied 23 absorption systems and concentrated mainly on the
Mg{\sc \,ii} $\lambda\lambda$2796/2803 doublet and the five strong
Fe{\sc \,ii} transitions redwards of 2340\,\AA. The single absorber
which dominates their sample, with a quoted uncertainty on
$\Delta\alpha/\alpha$ of just $0.1\times10^{-5}$, lies at redshift
$z_{\rm abs}=1.2433$. This places the three Fe{\sc \,ii} transitions
used in their analysis at observed wavelengths between $5250$ \&
$5340$\,\AA\ and the Mg{\sc \,ii} doublet appears at around
$6270$\,\AA. Fig.~\ref{fig:ESO} implies that a spurious velocity shift
of $75\,\ms$ between the Fe{\sc \,ii} and Mg{\sc \,ii} lines will be
measured in this system. The three Fe{\sc \,ii} lines all have large,
positive $q$-coefficients $\sim$1400${\rm \,cm}^{-1}$ while the Mg{\sc
  \,ii} doublet transitions both have very small $q$-coefficients
$\sim$150${\rm \,cm}^{-1}$. Equation (\ref{eq:da}) therefore implies
that the $75\,\ms$ velocity shift corresponds to a spurious
$(\Delta\alpha/\alpha)_{\rm ThAr}\approx +0.4\times10^{-5}$, four
times larger than the 1-$\sigma$ uncertainty quoted by
\citet{ChandH_04a}.

\begin{figure*}
\centerline{
 \hbox{
  \includegraphics[height=0.40\textwidth,angle=270]{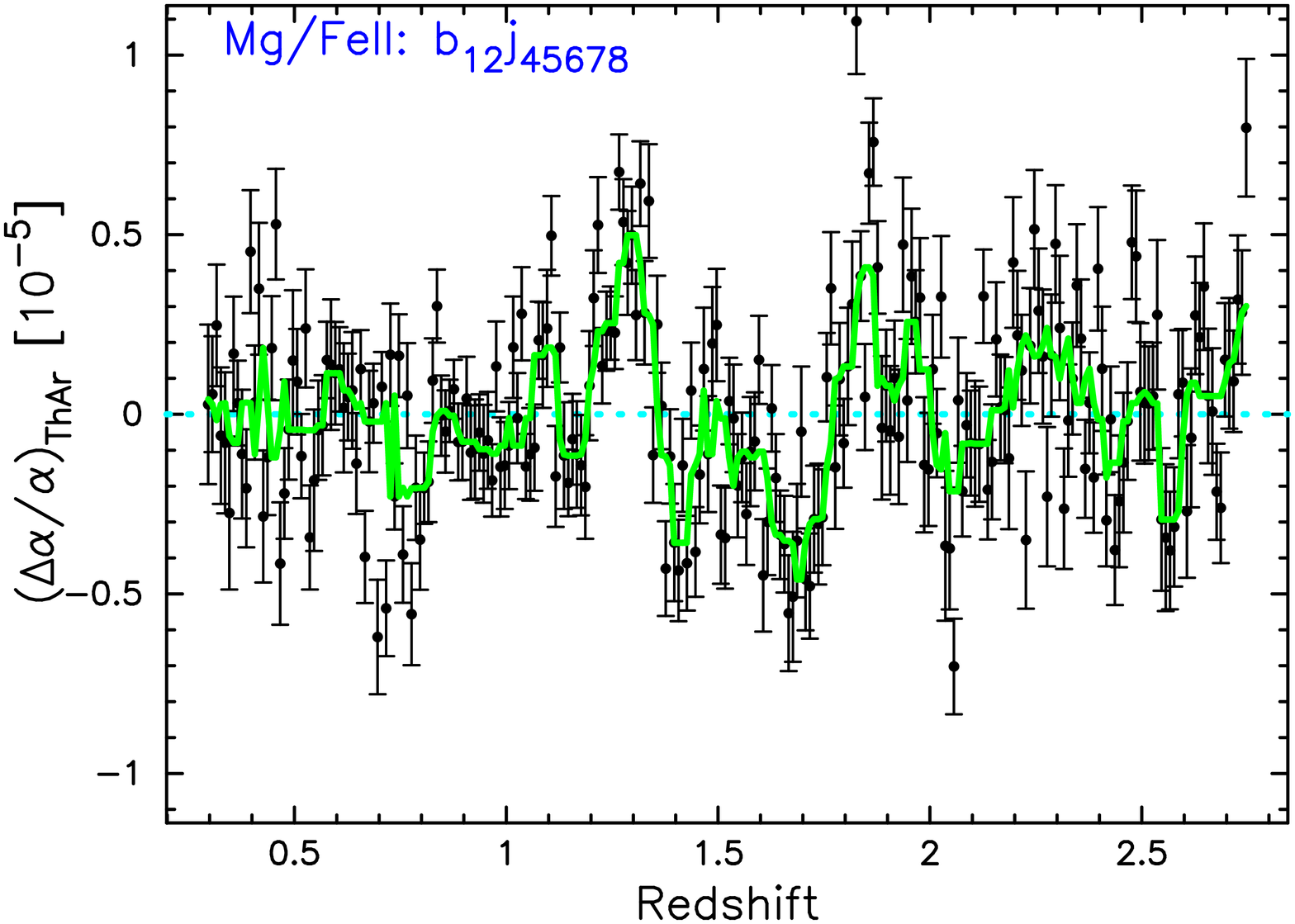}
  \hspace{0.01\textwidth}
  \includegraphics[height=0.40\textwidth,angle=270]{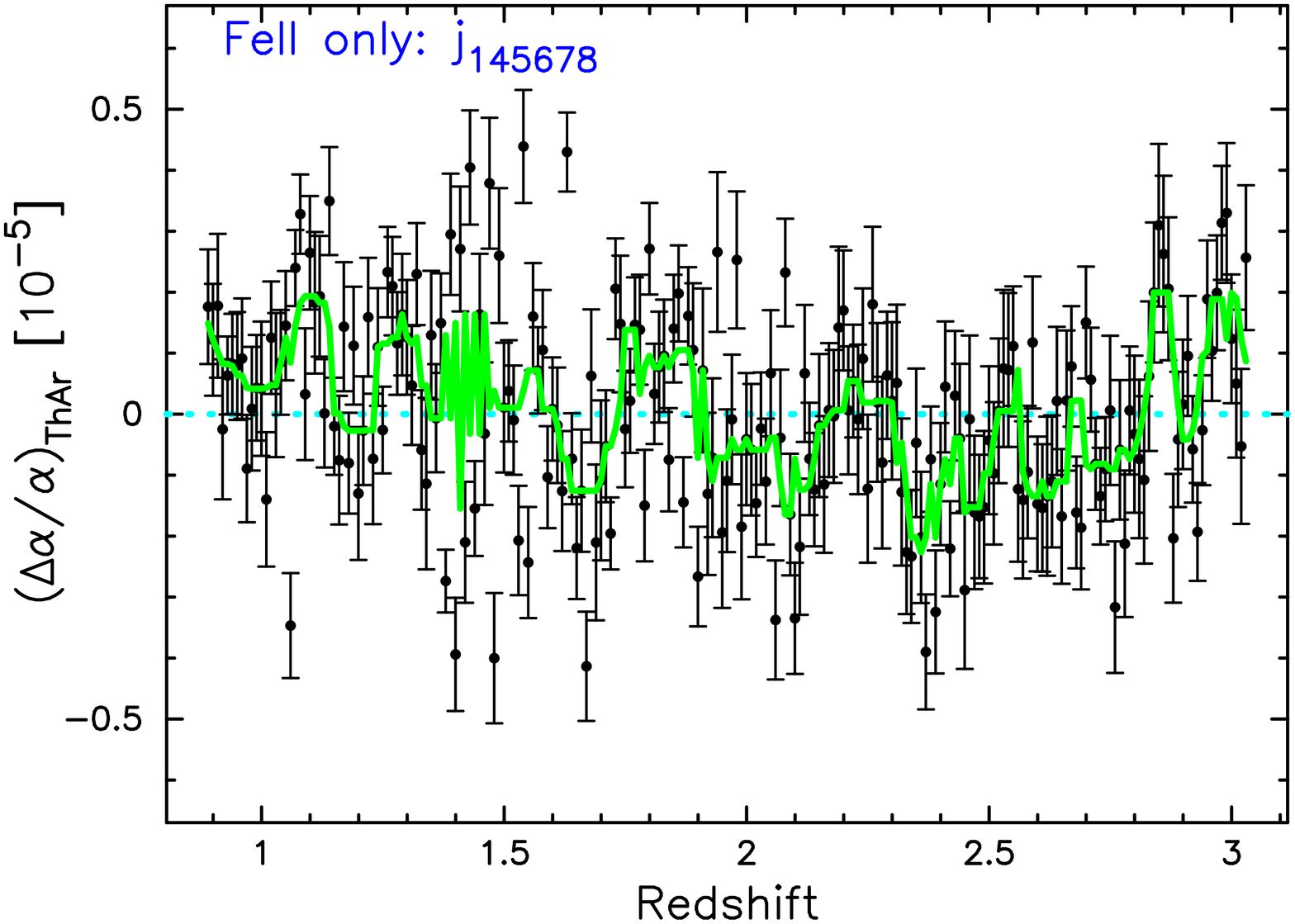}
 }
}
\caption{The spurious value of $\Delta\alpha/\alpha$ implied by the
  residuals in Fig.~\ref{fig:ESO}, $(\Delta\alpha/\alpha)_{\rm ThAr}$,
  as a function of redshift, assuming two different sets of
  transitions. The left panel utilises the Mg{\sc \,ii}
  $\lambda\lambda$2796/2803 doublet and Fe{\sc \,ii} $\lambda$2344,
  2374, 2382, 2586 \& 2600 lines. The right panel uses the same Fe{\sc
    \,ii} lines plus the Fe{\sc \,ii} $\lambda$1608 transition. The
  solid grey/green line shows a running median of
  $(\Delta\alpha/\alpha)_{\rm ThAr}$ over 7 redshift bins. The
  redshift bins are $\Delta z=0.01$ wide. If one had measured a value
  of $\Delta\alpha/\alpha$ from an absorber at a given redshift in a
  spectrum reduced with the unmodified UVES pipeline and calibrated
  using the ESO line-list, one would have to subtract the value of
  $(\Delta\alpha/\alpha)_{\rm ThAr}$ at that redshift to derive a more
  reliable estimate.}
\label{fig:fsc}
\end{figure*}

This example of converting the residuals in Fig.~\ref{fig:ESO} into
predictions of the systematic error in $\Delta\alpha/\alpha$ are
easily generalised to any absorption redshift once a set of $n\!>\!1$
transitions is specified. Fig.~\ref{fig:fsc} shows two examples, one
utilising the Mg{\sc \,ii} doublet and the five redder Fe{\sc \,ii}
transitions, the other using those five Fe{\sc \,ii} lines plus the
Fe{\sc \,ii}\,$\lambda$1608 transition. At each step in redshift, the
value of $(\Delta\alpha/\alpha)_{\rm ThAr}$ is calculated as follows:
for each transition $i$, the 7 ThAr lines which fall closest to its
redshifted wavelength are identified and their mean residual, $v_i$,
(as defined in Fig.~\ref{fig:ESO}) is calculated. The error in each
$v_i$ is assumed to be the same and is set equal to the RMS residual
of all $7\times n$ ThAr lines, divided by $\sqrt{7}$. A linear
least-squares fit between the $v_i$ values and the corresponding
$q$-coefficients [equation (\ref{eq:da})]\footnote{Although equation
  (\ref{eq:da}) formally has zero $y$-intercept, the fit between the
  $v_i$ and $q$ values must contain the $y$-intercept as a free
  parameter, to be determined simultaneously with
  $(\Delta\alpha/\alpha)_{\rm ThAr}$. This amounts to determining the
  best-fitted redshift of the set of the ThAr lines, mimicking the
  situation in the quasar spectra where one must determine the
  best-fitting $\Delta\alpha/\alpha$ and redshift simultaneously.}
then yields the best-fitted value of $(\Delta\alpha/\alpha)_{\rm
  ThAr}$ and its uncertainty.

\begin{figure}
\centerline{\includegraphics[width=1.00\columnwidth]{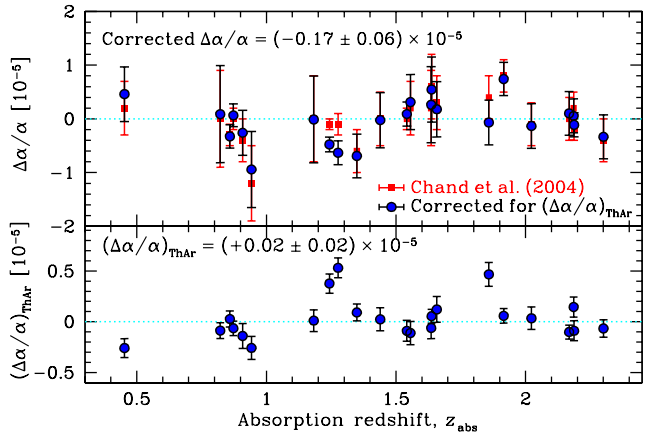}}
\caption{Effect of wavelength calibration errors on the
  $\Delta\alpha/\alpha$ values of \citet{ChandH_04a}. The lower panel
  gives the systematic error $(\Delta\alpha/\alpha)_{\rm ThAr}$ for
  each absorber while the upper panel shows the original
  $\Delta\alpha/\alpha$ values (red/grey squares) corrected by their
  respective $(\Delta\alpha/\alpha)_{\rm ThAr}$ values (solid
  circles). Even though the weighted mean calibration error is just
  $(\Delta\alpha/\alpha)_{\rm ThAr}=(+0.02\pm0.02)\times10^{-5}$, the
  fact that some of the (apparently) most precise
  $\Delta\alpha/\alpha$ values have large corrections means that the
  corrected weighted mean $\Delta\alpha/\alpha$ shifts significantly
  to $\Delta\alpha/\alpha=(-0.17\pm0.06)\times10^{-5}$.}
\label{fig:thar}
\end{figure}

The lower panel of Fig.~\ref{fig:thar} shows the values of
$(\Delta\alpha/\alpha)_{\rm ThAr}$ for each of the 23 absorbers in the
\citet{ChandH_04a} analysis. As described above, some absorbers are
particularly sensitive to miscalibrations due to their redshift and
the particular transitions used to determine $\Delta\alpha/\alpha$.
However, for 18 of the 23 absorbers, $\left|(\Delta\alpha/\alpha)_{\rm
    ThAr}\right|$ is less than $0.2\times10^{-5}$. It is nevertheless
important to asses the impact this has on the overall weighted mean
value of $\Delta\alpha/\alpha$. The values of $\Delta\alpha/\alpha$,
corrected by their respective $(\Delta\alpha/\alpha)_{\rm ThAr}$
values, are shown in the upper panel of Fig.~\ref{fig:thar}. The
errors bars are the quadrature addition of the original errors quoted
by \citet{ChandH_04a} and the errors in $(\Delta\alpha/\alpha)_{\rm
  ThAr}$ shown in the lower panel. The weighted mean of the corrected
values is $\Delta\alpha/\alpha=(-0.17\pm0.06)\times10^{-5}$; a
2.5-$\sigma$ `detection' of variation in $\alpha_{\rm em}$ compared to
the null result of $\Delta\alpha/\alpha=(-0.06\pm0.06)\times10^{-5}$
claimed by \citet{ChandH_04a}. Note that the error bars on the
$\Delta\alpha/\alpha$ values from \citet{ChandH_04a} plotted in
Fig.~11 have been shown to be significant underestimates in
\citet*{MurphyM_07d,MurphyM_07a}.  Thus, once larger, more appropriate
error bars are used, the relative importance of wavelength calibration
errors on the corrected \citet{ChandH_04a} result diminishes
\citep{MurphyM_07d}.  Nevertheless, Figs.~\ref{fig:fsc} \&
\ref{fig:thar} clearly demonstrate how the ESO line-list and the
wavelength calibration procedure of the original UVES pipeline cause
significant systematic effects in measurements of
$\Delta\alpha/\alpha$ at the $0.1\times10^{-5}$ precision level.

A similar analysis is presented by \citet{ChandH_06a} who
cross-correlate two UVES ThAr spectra, one calibrated using the ESO
line-list and the other calibrated using the line-list distributed by
the National Optical Astronomy
Observatory\footnote{http://www.noao.edu/kpno/specatlas/thar/thar.html}
(NOAO), to determine the distortions of the wavelength scale over the
range $3100$--$6700$\,\AA. They find very similar structures (in their
figure 7) to those seen in Fig.~\ref{fig:ESO}, including the sharp
spike around $6300$\,\AA. \citet{ChandH_06a} then test the effect
these distortions would have on a single Mg/Fe{\sc \,ii} absorber at
$z_{\rm abs}=1.15$, using only the redder Fe{\sc \,ii} lines and the
Mg{\sc \,ii} doublet\footnote{\citet{ChandH_06a} do not consider the
  Fe{\sc \,ii}\,$\lambda$1608 line in this test despite the fact that
  their analysis of the real absorber relies most heavily on this
  single transition.}. Since these lines are all redshifted to
wavelengths where the mean calibration residual is similar
(i.e.~between $5040$ \& $6030$\,\AA) -- falling just short of the
spike around $6300$\,\AA\ -- it is not surprising to find little-to-no
effect on $\Delta\alpha/\alpha$ at this redshift. Based on this
single-redshift example, \citet{ChandH_06a} claim that current UVES
constraints on $\Delta\alpha/\alpha$ should be fairly immune to any
problems introduced by the ESO line-list and (unmodified) UVES
pipeline wavelength calibration procedures. As demonstrated in the
specific examples above and in the more general cases illustrated in
Figs.~\ref{fig:fsc} \& \ref{fig:thar}, such a simplistic extrapolation
to other redshifts significantly underestimates the systematic errors
involved.

Finally, note that in \citet{MurphyM_01b} and \citet{MurphyM_03a} we
carried out a similar analysis on the Keck/HIRES spectra which
provided our previous constraints on $\Delta\alpha/\alpha$.  The
wavelength calibration employed the NOAO ThAr line-list discussed
above. In those works we found no such structures like those seen in
Fig.~\ref{fig:fsc}. Thus, although the current UVES constraints on
$\Delta\alpha/\alpha$ have been adversely affected by the ESO
line-list, such a problem does not generally exist for HIRES
constraints to any significant degree.

\section{Conclusions}\label{sec:conc}

We have implemented a simple iterative algorithm to select ThAr lines
for use in calibrating echelle spectrographs. The algorithm combines
the expected line properties (e.g.~degree of blending), based on the
input line-list, with properties of the ThAr spectrum observed on the
spectrograph of interest (e.g.~resolution, relative line intensities
etc.). We emphasise that although this paper has focused on the
application of this algorithm to VLT/UVES spectra, the algorithm can
be applied to any other spectrograph; all that is required is a ThAr
spectrum taken on that spectrograph and a reduction code which allows
repeatable wavelength calibration with different ThAr line-lists.

Using the most recent and reliable ThAr laboratory wavelengths,
including that of \citet{LovisC_07a}, we have produced a new ThAr
line-list which, we argue, should be used to calibrate UVES spectra in
preference to the list distributed by ESO. The main advantages of the
former over the latter are: (i) more homogeneous input line catalogues
with mutually consistent wavelength scales; (ii) more accurate and
precise laboratory wavelengths; (iii) only unblended lines appear in
the final list; and (iv) only lines which are {\it observed} to be useful
in UVES spectra are included. Indeed, the wavelength calibration
residuals achieved using the new UVES line-list (${\rm
  RMS}\!\sim\!35\,\ms$) are more than a factor of 3 better than those
achieved with the ESO or \citet{DeCuyperJ-P_98a} list (${\rm
  RMS}\!\sim\!130\,\ms$) -- see Fig.~\ref{fig:wavres}. An assumption one
makes in using our new line-list to calibrate UVES spectra is that the
relative line intensities (i.e.~lamp conditions like pressure, current
etc.) are similar to those in the sample UVES ThAr spectra used in our
selection algorithm. However, ideally, the selection algorithm should
be included in the wavelength calibration procedure itself, thus
producing the best selection of ThAr lines for each individual ThAr
exposure being reduced.

The new line-list has been used to probe distortions of the wavelength
scale introduced into UVES spectra which were originally wavelength
calibrated with the ESO line-list. The distortions include sharp,
short-range corrugations with peak-to-peak amplitude up to
$\sim$75\,\ms\ and longer-range variations between $+30\,\ms$ and
$-30\,\ms$ (Fig.~\ref{fig:ESO}). These distortions have serious
implications in the context of current UVES estimates of possible
variations in the fine-structure constant, especially those of
\citet{ChandH_04a} -- see Figs.~\ref{fig:fsc} \& \ref{fig:thar}. For
example, if one employs the common Mg/Fe{\sc \,ii} combination of
transitions to derive $\Delta\alpha/\alpha$ then the errors in the ESO
line-list cause a spurious value of $(\Delta\alpha/\alpha)_{\rm
  ThAr}\approx+0.4\times10^{-5}$ for absorption redshifts
$1.22\!\la\!z_{\rm abs}\!\la\!1.35$ and $(\Delta\alpha/\alpha)_{\rm
  ThAr}\approx-0.3\times10^{-5}$ for $1.38\!\la\!z_{\rm
  abs}\!\la\!1.75$.  Although \citet{ChandH_04a} claim an overall
precision in $\Delta\alpha/\alpha$ of $0.06\times10^{-5}$ from 23
Mg/Fe{\sc \,ii} absorbers \citep[but see][]{MurphyM_07d,MurphyM_07a},
a single absorber at $z_{\rm abs}=1.2433$ strongly dominates their
sample with an individual error on $\Delta\alpha/\alpha$ of
$0.1\times10^{-5}$.  However, at this redshift, and considering which
transitions are used to derive $\Delta\alpha/\alpha$ in this system,
we find that a spurious systematic shift of
$(\Delta\alpha/\alpha)_{\rm ThAr}\approx+0.4\times10^{-5}$ has
affected this system. Once such wavelength calibration corrections are
applied to the entire sample, the null result of
$\Delta\alpha/\alpha=(-0.06\pm0.06)\times10^{-5}$ claimed by
\citet{ChandH_04a} becomes
$\Delta\alpha/\alpha=(-0.17\pm0.06)\times10^{-5}$ -- a 2.5-$\sigma$
`detection'. The new ThAr line-list we present here should aid in
avoiding such important yet elementary errors in future analyses of
UVES spectra.

\section*{Acknowledgments}

We thank R.~Engleman and W.~Whaling for detailed discussions about the
Th and Ar line-lists, respectively, and for providing the lists in
electronic form. R.~Fear of Cathodeon, manufacturer of ThAr hollow
cathode lamps, also provided helpful information concerning the
operating conditions of such lamps. MTM thanks PPARC for an Advanced
Fellowship at the IoA.


\section*{Supplementary material}

The following supplementary material is available for this article
online on {\it Synergy} and at
http://www.ast.cam.ac.uk/$\sim$mim/pub.html.

{\bf Table \ref{tab:list}.} The final ThAr line-list for UVES. In
addition to an ASCII version, we also provide a version in a format
which is readable by the UVES pipeline distributed by ESO.

\bspsmall

\label{lastpage}

\end{document}